\documentclass[12pt]{article} 
\usepackage{array,epsfig,fancyheadings,rotating}
\usepackage[]{hyperref} 

\usepackage{amsmath}
\usepackage{amssymb}
\usepackage{amsfonts}
\usepackage{multirow}
\usepackage{amsthm}

\newtheorem{theorem}{Theorem}

\newtheorem{corollary}{Corollary}

\theoremstyle{definition}

\title{A Hybrid Symbolic/Numeric Solution  To Polynomial SEM}
\author{Reinhard Oldenburg, Augsburg University, Augsburg, Germany}
\date{10th September 2021}

\begin{document}

\maketitle

\begin{abstract}
	There are many approaches to nonlinear SEM (structural equation modeling)
	but it seems that a rather straightforward approach using Isserlis' theorem has not yet been investigated although it
	allows the direct extension of the standard linear approach to nonlinear linear SEM. The reason may be that this method requires some symbolic
	calculations done at runtime. This paper describes the class of appropriate models and outlines the algorithm that calculates the covariance matrix and higher moments. Simulation studies show that the method works very well and especially that tricky models can be estimated accurately by taking higher movements into account, too. 
	
\end{abstract}

\def\thefigure{\arabic{figure}}
\def\thetable{\arabic{table}}

\renewcommand{\theequation}{\thesection.\arabic{equation}}

\section{ Introduction}
Linear SEM is a standard statistical method in the social sciences and recently interest in nonlinear SEM emerged for many reasons: Interaction and nonlinear effects are crucial in many applications in psychology. An early overview that discusses many foundational issues is given by \cite{SM1998}. For interaction effects of manifest variables the product indicator approach \cite[ p. 441]{Hoyle2012} is an obvious approach and there are several techniques to handle the constraints it imposes on the parameters to be estimated. More flexible are approaches that analyze the distributional consequences of the nonlinear relations -- see \cite{KWS2011} and \cite{UNBK2017} for an overview. However, many of these approaches are restricted to quadratic relations. A overview of various methods suitable for the quadratic case is given in \cite{HWH2012}.  Beside Bayesian estimation they find the LMS method of \cite{KM2000} to perform quite well. 
While quadratic models are sufficient for many practical situations it is both of theoretical as well practical (e.g. interactions of three variables) interest to allow general polynomial relations and the present paper presents an approach that provides this possibility. Full source code of the new method is available.

\section{Model class and algorithm}
The class of models that this paper deals with consists of models that separate both latent and manifest variables into two groups. Like \cite[pp. 319]{Bollen1989} I assume that there are measurement models for
 $k$  exogenous ($\xi$) and  $l$  endogenous ($\eta$) latent variables in terms of  $m=m_1+m_2$  observed variables  $x,y$: 

\begin{equation}\label{2.1}
	x=\Lambda_x\xi +\delta ,y=\Lambda_y\eta +\epsilon 
\end{equation}
and that  $\xi ,x,\delta $ \ are jointly normally distributed with zero expectation (this last assumption is not severe of course as the mean structure is rather trivial and the assumption can be be fulfilled
by subtracting the means of observed data). No normality assumption is made for  $y$, of course, as nonlinear relations imply other distributions. The structural model is given by a polynomial function and it involves another (vector) error term  $\zeta $: 

\begin{equation}\label{2.2}
	\eta =f(\xi )+\zeta  
\end{equation}

Note that  $\xi $  and  $\eta $  are vectors of random variables and thus  $f:\mathbb{R}^k\rightarrow \mathbb{R}^l$  consists of  $l$  real multivariate polynomial functions. Thus, it can be written as

\begin{equation}\label{2.3}
	f\left(\xi \right)=\sum_{\left(e_1,{\dots},e_k\right){\in}\mathbb{N}_0^k}c_{\left(e_1,{\dots},e_k\right)}{\cdot}\xi
	_1^{e_1}{\cdot}{\dots}{\cdot}\xi_k^{e_k}
\end{equation}
where only finitely many of the  $c_{\left(e_1,{\dots},e_k\right)}{\in}R^l$ \ are nonzero.

All latent variables will be assumed to have zero expectation. If this is not the case, they can be replaced by the sum of a scalar parameter and a new, centered latent variable.

Furthermore, it is assumed that all components of error vectors are independent of each other and moreover 
\begin{equation}\label{2.4}
	\mathit{cov}\left(\delta ,\epsilon \right)=\mathit{cov}\left(\xi ,\delta \right)=\mathit{cov}\left(\xi ,\epsilon
	\right)=\mathit{cov}\left(\zeta ,\delta \right)=\mathit{cov}\left(\zeta ,\epsilon \right)=0
\end{equation}

As in the linear case these independency assumptions can be relaxed somewhat by allowing some covariance to be non-zero, but this is limited by the identification problem, of course. 

Now the model class is specified. Before presenting the algorithm, we need to recall the technical theorem that it is based on:

\begin{theorem} (\cite{I1918}) Assume  $X_1,{\dots},X_n$ to be multivariate normally distributed and centered (i.e. expectation  $E\left(X_i\right)=0$) then the expectation of their product can be expressed in terms of covariances:
\begin{equation}\label{2.5}
E\left(X_1{\cdot}{\dots}{\cdot}X_n\right)=\sum_{p{\in}P_n^2}\prod
	_{\left\{i,j\right\}{\in}p}\mathit{cov}(X_i,X_j)
\end{equation}

Where  $P_n^2$ is the set of all partition of   $\{1,{\dots},n\}$  into disjoint subsets of size 2.
\end{theorem}

Under the assumptions given above we will present a straightforward algorithm to calculate the implied covariance matrix   
\begin{equation}\label{2.6}
	\Sigma
	=\left(\begin{matrix}\mathit{cov}(x,x')&\mathit{cov}(y,x')\\\mathit{cov}(y,x')&\mathit{cov}(y,y')\end{matrix}\right)
\end{equation}

The first entry is calculated exactly as in \cite[pp. 323]{Bollen1989}:
\begin{equation}\label{2.7}
	\begin{split}
	\mathit{cov}\left(x,x'\right)=E\left(xx'\right)=E\left(\left(\Lambda_x\xi +\delta \right)\left(\xi '\Lambda
	_x'+\delta '\right)\right)=\\
	\Lambda_x E\left(\mathit{\xi \xi }'\right)\Lambda_x'+\Lambda_xE\left(\xi \delta' \right)+E\left(\delta \xi'\right)\Lambda_x'+E\left(\delta \delta '\right)=\Lambda_xE\left(\mathit{\xi \xi
	}'\right)\Lambda_x'+E\left(\delta \delta'\right)
	\end{split}
\end{equation}
Here, the last summand is a diagonal matrix because of the independency assumptions made above. Now, turn to the off-diagonal entry
\begin{equation}\label{2.8}
	\begin{split}	\mathit{cov}\left(y,x'\right)=E\left(\left(\Lambda_y(f(\xi )+\zeta )+\epsilon \right){\cdot}\left(\xi '\Lambda
	_x'+\delta '\right)\right)=\\
	E\left(\left(\Lambda_y(f(\xi )+\zeta )\right){\cdot}\xi '\Lambda
	_x'\right)=E\left(\Lambda_yf(\xi ){\cdot}\xi '\Lambda_x'\right)=\Lambda_yE\left(f(\xi ){\cdot}\xi '\right)\Lambda
	_x'=\\
	\sum_{\left(e_1,{\dots},e_k\right){\in}N_0^k}\Lambda_y{\cdot}c_{\left(e_1,{\dots},e_k\right)}{\cdot}E(\xi
	_1^{e_1}{\cdot}{\dots}{\cdot}\xi_k^{e_k}{\cdot}\xi ')\Lambda_x'
		\end{split}
\end{equation}
The last entry is not yet fully calculated, but it is clear that one needs only to evaluate the expectation on monomials of centered, normally distributed variables and therefore Isserlis' theorem can be applied so that the result is a polynomial in covariances and parameters.
\begin{equation}\label{2.9}
		\begin{split}
	\mathit{cov}\left(y,y'\right)=&\\
	&E\left(\left(\Lambda_y\left(f\left(\xi \right)+\zeta \right)+\epsilon
	\right){\cdot}\left(\left(f\left(\xi \right)'+\zeta '\right)\Lambda_y'+\epsilon '\right)\right)-\\
	&E\left(\Lambda
	_y\left(f\left(\xi \right)+\zeta \right)+\epsilon \right){\cdot}E\left(\left(f\left(\xi \right)'+\zeta
	'\right)\Lambda_y'+\epsilon '\right)=\\
	&E\left(\left(\Lambda_y\left(f\left(\xi \right)+\zeta
	\right)\right){\cdot}\left(f\left(\xi \right)'+\zeta '\right)\Lambda_y'\right)+E\left(\epsilon \epsilon
	'\right)=\\
	&\Lambda_yE\left(f\left(\xi \right)f\left(\xi \right)'+f\left(\xi \right)\zeta '+\mathit{\zeta f}\left(\xi
	\right)'+\zeta \zeta '\right)\Lambda_y'+E\left(\epsilon \epsilon '\right)=\\
	&\Lambda_yE\left(f\left(\xi
	\right)f\left(\xi \right)'+f\left(\xi \right)\zeta '+\mathit{\zeta f}\left(\xi \right)'\right)\Lambda_y'+\Lambda
	_yE\left(\mathit{\zeta \zeta }'\right)\Lambda_y'+E(\epsilon \epsilon ')
	\end{split}
	\end{equation}

Again, this is not yet fully calculated but it is obvious that linearity of the expectation and the polynomial structure of  $f$  allows this to be expanded so that Isserlis' theorem can be applied. 
Collecting the above results, one arrives at:

\begin{theorem} The model-based covariance matrix  $\Sigma $  of the polynomial SEM defined above consists of polynomials in the parameters of  $\Lambda_x,\Lambda_y$  as well as the variances of  $\epsilon ,\zeta ,\delta $ and variances and covariances of  $\xi $. 
\end{theorem}

Inspecting the logic of the calculations done in the proof reveal the following generalization:
\begin{corollary}
	Any moment $E(\prod_{i=1}^{m_1}x_i^{k_i}\prod_{i=1}^{m_2}y_i^{l_i})$ of the polynomial SEM defined above can be expressed as polynomials in the parameters of  $\Lambda_x,\Lambda_y$  as well as the variances of  $\epsilon ,\zeta ,\delta $ and variances and covariances of  $\xi $. 
\end{corollary}

This corollary gives the possibility to use information that is contained in higher order moments of the manifest variables.

The difference to linear SEM is that the entries in  $\Sigma$ are polynomials of higher degree. Now, in principle any estimation method that minimizes some distance measure between  $\Sigma$  and  the covariance matrix $S$  of the data can be applied. However, for nonlinear  $f$  and normal  $x$  it is clear that  $y$  will not be normally distributed. Hence,  \begin{equation}\label{2.10}
	F_{\mathit{ULS}}=\frac 1 2\mathit{tr}(\left(S-\Sigma \right)^2)
\end{equation}  
is a good choice as a discrepancy function as it does not depend on distributional assumptions. In contrast  
$F_{\mathit{ML}}=\mathit{tr}\left(S\Sigma ^{-1}\right)+\log \left|\Sigma
\right|-\log \left|S\right|-m$  
will not lead to consistent estimations because of  $y$  violating normality
assumptions. However, the  $x$  part of the data is required to be multivariate normal and thus the following mixed strategy is obvious: The blocks of  $\Sigma
=\left(\begin{matrix}\mathit{cov}(x,x')&\mathit{cov}(y,x')\\\mathit{cov}(y,x')&\mathit{cov}(y,y')\end{matrix}\right)$
 are estimated with different methods: the whole objective function will be
\begin{equation}\label{2.11}F_{\mathit{ML}}\left(\mathit{cov}\left(x,x'\right)\right)+2F_{\mathit{ULS}}\left(\mathit{cov}\left(y,x'\right)\right)+F_{\mathit{ULS}}(\mathit{cov}\left(y,y'\right))
\end{equation}
However, other methods provide even better estimates and thus we will not investigate this path further.

A sound theoretical basis has WLS estimation (weighted least square) based on the theory developed by \cite{Browne}, see also \cite[p. 426]{Bollen1989}. Note, that what is mostly denoted by GLS is a special case of this theory for normal data but, of course, in the present case it is crucial to implement the general case. For the reader's convenience this approach is recalled here:
The objective function in terms of the collected  parameter vector  $\theta $  is  \begin{equation}\label{2.12}F_{\mathit{WLS}}\left(\theta
\right)=\left(s-\sigma \left(\theta \right)\right)'{\cdot}W^{-1}{\cdot}(s-\sigma \left(\theta \right))
\end{equation} 
where 
$s,\sigma$  are vector versions of  $S,\Sigma$, i.e. with the diagonal and upper entries of the covariance matrices flattened out in vector form, and  $W$  is a weight matrix. Browne's strong result is that estimation by  $F_{\mathit{WLS}}$  where  $W$   is chosen to be the difference between that matrix of 4\textsuperscript{th} order moments and the products of covariances (see \cite[eq. (3.4)]{Browne} then one gets asymptotically distribution-free estimation. 

As noted above the calculations strategy allows to investigate higher moments as well. Denote the generalizations of covariance by
\begin{equation}\label{2.13}
	\mathrm{cov}^{(k)}(X_{i_1},\ldots,X_{i_k}):=E(X_{i_1}\cdot\ldots\cdot X_{i_k})-E(X_{i_1})\cdot\ldots\cdot E(X_{i_k}), k\in\mathbb{N}
\end{equation} 
Combining the manifest variables into a single vector $z=(x,y)$ of $m$ random variables. Then one may calculate the theoretical moments $\Sigma^{(k)}_{i_1,\ldots,i_{k'}}:=	\mathrm{cov}^{(k)}(z_{i_1},\ldots,z_{i_k})$  by inserting the formulas \ref{2.1} and the corresponding empirical momements from the data $S_{i_1,\ldots,i_k}:=	\mathrm{cov}^{(k)}(z_{i_1},\ldots,z_{i_k})$ by inserting the data.  Then it is natural to define 
\begin{equation}\label{2.14}
	F_{ULS^{(k)}}:=\sum_{k'=2}^{k} \frac{1}{2 ^{k'-1} }    \sum_{i_1=1,\ldots,i_{k'}=1}^{m,\ldots,m}    \left(\Sigma^{(k')}_{i_1,\ldots,i_{k'}}   - S_{i_1,\ldots,i_{k'}} \right)^2
\end{equation} 
Then $F_{ULS}=F_{ULS^{(2)}}$. In simulations studies below $F_{ULS^{(3)}}$ will be used and turns out to give superior performance.

Using symbolic computation one can implement this method rather easily: The model equations are specified as replacement rules and applied to the symbolic (higher order) covariance matrix. Then linearity of $cov$ can be applied until the above calculations using Isserlis' theorem can be performed. Finally, the objective functions can be put together rather easily. Special care has to be applied in performing the numeric optimization that estimates the parameters. There is a risk of getting stuck with only a local minimum and e.g. for GLS it is advisable to take ULS estimated as initial values. 

The implementation has been done in Mathematica and the code is public \cite{Old2020}. Especially the simulations presented in this paper (including full source code) are available at \url{https://myweb.rz.uni-augsburg.de/~oldenbre/sem/Polysem.pdf} (PDF) and \url{https://myweb.rz.uni-augsburg.de/~oldenbre/sem/Polysem.nb} (Mathematica notebook). Furthermore, there is an implementation in R \url{https://myweb.rz.uni-augsburg.de/~oldenbre/sem/polysem.R} that performs the same calculations but is much slower and less accurate. The implementation  in Mathematica is faster but still the estimation of one sample for Ganzach's model with $n=1000$ takes about four minutes on a modestly fast two-core notebook computer with 8 GB RAM.

\section{Case studies}
This paragraph reports on results of a practical application of this algorithms. The polynomial methods described in this paper are carried out with four of the estimation methods described above. Computations were done in the Mathematica system. 

As first test model I use the  quadratic model given by  \cite{AM2021}. This model has four latent variables $\eta_1,\ldots,\eta_4$ each measured by three manifest variables $y_1,\ldots,y_{12}$. The structural model is: $\eta_3=B_1\eta_1 + B_2\eta_2 + B_3\eta_1 \cdot\eta_2+\epsilon_3,\eta_4=B_2\eta_3+\epsilon_4$.

The generation of the sample data sets is done by the following algorithm that tries to mimic the data generation in their publication:

\begin{enumerate}   
	\item $\eta_1,\eta_2$ are sampled normally distributed with mean 0 and covariance matrix $\left(\begin{array}{cc}
		1.2 & 0.4 \\
		0.4 & 0.8
	\end{array}\right)$
	\item $\eta_3:=B_1\cdot\eta_1 + B_2\cdot\eta_2 + B_3\cdot\eta_1 \cdot\eta_2+N(0,0.2,n),
	\eta_4:=B_4\cdot\eta_3+N(0,0.1,n) $ with $B_1 = 0.1,  B_2 = 0.3, B_3 = 0.2, B_4 = 0.7$
	\item $i\in\{1,..,12\}: y_i:=c_i \cdot\eta_{\lceil i/3\rceil}+N(0,0.1\cdot(1+(i \mod 3)),n)$ with $c=(1,0.5,0.7,1,0.7,0.4,1,1.2,0.4,1,0.8,0.9)$
\end{enumerate}

For the sake of comparing different methods 100 samples of data sets with $n=1000$ were created. Table 1 shows the mean errors and the standard deviations (in parentheses) of the estimates for the four central path weights $B_1,...,B_4$.  The other parameters are equally well estimated by all methods. The conclusion will be given after presenting the second example.

\begin{table}[t!] 
	\caption{Results of simulation study for the first test model}
	\label{tab:simulation}\par
	\resizebox{0.9\linewidth}{!}{\begin{tabular}{|r|rrrr|} \hline 
			Variable & ULS & ULS3 & GLS & FWLS \\ \hline
			$B_1$ & -0.006(0.007) & -0.006(0.007) & -0.005(0.008) & -0.003(0.008) \\  
			$B_2$ & -0.008(0.019) & -0.006(0.020) & -0.007(0.019) & -0.005(0.019)\\
			$B_3$ & -0.206(0.008) & 0.005(0.015) & 0.114(0.004) & -0.206(0.008)\\
			$B_4$ & -0.005(0.015) & -0.007(0.012) & -0.008(0.011) & -0.008(0.012)	\\ \hline
	\end{tabular}}
\end{table}

A second, more demanding test case is Ganzach's model as studied in \cite{KB2009}. This model has three latent variables $\eta,\xi_1,\xi_2$ each measured by three manifest variables $i=1...3: x_i=\lambda_i\xi_1+\delta_i, i=4...6: x_i=\lambda_i\xi_2+\delta_i, y_i=\mu_i\eta+\epsilon_i, \lambda_1=\lambda_4=\mu_1=1$.  
The structural model is: $\eta=\gamma_1\xi_1+\gamma_2\xi_2+\omega_{11}\xi_1^2+\omega_{22}\xi_2^2+\omega_{12}\xi_1\xi_2+\epsilon_0$.

The simulated data for this model were generated by the following algorithm: 1000 cases were sampled according to the following algorithm (where  $N(\mu ,\sigma )$  denote normally distributed random values):

\begin{enumerate}
	\item $\xi_1,\xi_2$ \ are sampled normally distributed with mean 0 and covariance matrix
	$\left(\begin{matrix}1&0.2\\0.2&1\end{matrix}\right)$ 
	\item $\eta :=\gamma_1\xi_1+\gamma_2\xi_2+\omega_{11}\xi_1^2+\omega_{12}\xi_1\xi_2+\omega_{22}\xi_2^2+N(0,0.3)$
	 with $\gamma_1=0.3,\gamma_2=0.5,\omega_{11}=0.2,\omega_{12}=0.4,\omega_{22}=0.7$
	\item $i{\in}\left\{1,2,3\right\}:y_i:=d_i\eta +N\left(0,0.1\right),x_i:=c_i\xi
	_1+N\left(0,0.1\right),i{\in}\left\{4,5,6\right\}:x_i:=c_i\xi_2+N(0,0.1)$ \ with 
	$c=\left(1,0.7,1.2,1,0.5,0.9\right),d=(1,0.8,1.3)$
\end{enumerate}

The results for this study are given in table 2. All methods easily give correct estimations for  $\lambda_{ij}^x,\lambda_{ij}^y$, hence these are omitted from the following table. Instead, only he central weights with true values $\gamma_1=0.3,\gamma_2=0.5, \omega_{11}=0.2, \omega_{12}=0.4, \omega_{22}=0.7$ are investigated. The entries in the table give mean (over 100 samplings) of the differences between these true values and the estimates obtained. Standard deviations are given in parentheses. For nlsem (with method qml) only a small number of calculations was performed due to long run times and hence no standard deviations are given. 

\begin{table}[t!] 
	\caption{Results of simulation study for Ganzach's model}
	\label{tab:simulation}\par
	\resizebox{0.9\linewidth}{!}
	{\begin{tabular}{|r|rrrrr|} \hline 
		Variable & ULS & ULS3 & GLS & FWLS & nlsem \\\hline
$\gamma_1$ & -0.004(0.062) & -0.006(0.087) & -0.003(0.062) & -0.001(0.061)  & -0.021\\
$\gamma_2$ & -0.005(0.050) & -0.002(0.078) & -0.004(0.050) & -0.001(0.051) & -0.009\\
$\omega_{11}$ & -0.212(0.022) & -0.014(0.051) & -0.301(0.008) & -0.220(0.022) & 0.120\\
$\omega_{12}$ & -0.002(0.032) & -0.019(0.063) & -0.114(0.011) & -0.008(0.030) & 0.073 \\
$\omega_{22}$ & -0.118(0.022) & -0.006(0.051) & -0.258(0.041) & -0.126(0.022)	& - 0.019\\ \hline
	\end{tabular}}
\end{table}

Results for both models show that ULS3 performs much better than the other methods. Especially, the data suggest that the method may give unbiased estimations. This hypotheses is even further supported by the simulations with samples of size $n=10.000$ where error is even smaller. Especially note that nlsem \cite{UNBK2017} performs much worse than ULS3. 

\section{Conclusion}
The method presented in this paper is quite general as it can handle all polynomial SEM and yet the quality of estimations is very good. Two innovations are essential for this result: First, the use of symbolic computation at run-time to apply Isserlis' theorem, and second the fitting of higher moments that can be done on this basis.  As both Mathematica and R implementations are available, further research might now investigate the hypothesis that results are unbiased if the order of moments taken into account is sufficiently high.


\begin{thebibliography}{10}
	
	\bibitem{SM1998}
	R.~E. Schumacker and G.~A. Marcoulides.
	\newblock {\em Interaction and nonlinear effects in structural equation
		modeling}.
	\newblock Lawrence Erlbaum Associates, Mahwah, NJ, 1998.
	
	\bibitem{Hoyle2012}
	R.~H.~(Ed.) Hoyle.
	\newblock {\em Handbook of Structural Equation Modeling}.
	\newblock The Guilford Press, New York, 2012.
	
	\bibitem{KWS2011}
	A.~Kelava, Ch.~S Werner, K.~Schermelleh-Engel, H.~Moosbrugger, D.~Zapf, Y.~Ma,
	H.~Cham, L.~S. Aiken, and S.~G.. West.
	\newblock Advanced nonlinear latent variable modeling: Distribution analytic
	lms and qml estimators of interaction and quadratic effects.
	\newblock {\em Structural Equation Modeling}, 18:465--491, 2011.
	
	\bibitem{UNBK2017}
	N.~Umbach, K.~Naumann, H.~Bradt, and A.~Kelava.
	\newblock Fitting nonlinear structural equation models in r with package nlsem.
	\newblock {\em Journal of Statistical Software}, 77:7, 2017.
	
	\bibitem{HWH2012}
	J.~R. Harring, B.~A. Weiss, and J.~C. Hsu.
	\newblock A comparison of methods for estimating quadratic effects in nonlinear
	structural equation models.
	\newblock {\em Psychological Methods}, 17:193--214, 2012.
	
	\bibitem{KM2000}
	A.~Klein and H.~Moosbrugger.
	\newblock Maximum likelihood estimation of latent interaction effects with the
	lms method.
	\newblock {\em Psychometrika}, 65:457--474, 2000.
	
	\bibitem{Bollen1989}
	K.~A. Bollen.
	\newblock {\em Structural Equations with Latent Variables}.
	\newblock John Wiley, Hoboken, 1989.
	
	\bibitem{I1918}
	L~Isserlis.
	\newblock On a formula for the product-moment coeﬃcient of any order of a
	normal frequency distribution in any number of variables.
	\newblock {\em Biometrika}, 1-2(12):134--139, 1989.
	
	\bibitem{Browne}
	M.~W. Browne.
	\newblock Asymptotically distribution-free methods for the analysis of
	covariance structure.
	\newblock {\em British Journal of Mathematical and Statistical Psychology},
	37:62--83, 1984.
	
	\bibitem{Old2020}
	R.~Oldenburg.
	\newblock Structural equation modeling -- comparing two approaches.
	\newblock {\em The Mathematica Journal}, 22, 2020.
	
	\bibitem{AM2021}
	T.~Asparouhov and B.~Muthén.
	\newblock Bayesian estimation of single and multilevel models with latent
	variable interactions.
	\newblock {\em Structural Equation Modeling: A Multidisciplinary Journal},
	28(2):314--328, 2021.
	
	\bibitem{KB2009}
	A.~Kelava and H.~Brandt.
	\newblock Estimation of nonlinear latent structural equation models using the
	extended unconstrained approach.
	\newblock {\em Review of Psychology}, 16(2):123--131, 2009.
	
\end{thebibliography}
\end{document}